\documentclass[conference]{IEEEtran}
\IEEEoverridecommandlockouts

\usepackage{cite}
\usepackage{amsmath,amssymb,amsfonts}
\usepackage{graphicx}
\usepackage{optidef}
\usepackage{textcomp}
\usepackage{xcolor}
\usepackage{multirow}
\usepackage{algorithm}
\usepackage{algpseudocode}
\usepackage{float}
\usepackage{booktabs} 
\usepackage{setspace}

\usepackage{siunitx} 
\usepackage{subfig}
\usepackage[draft=true]{hyperref}
\hypersetup{bookmarks=false}
\graphicspath{ {./figures/} }
\def\BibTeX{{\rm B\kern-.05em{\sc i\kern-.025em b}\kern-.08em
    T\kern-.1667em\lower.7ex\hbox{E}\kern-.125emX}}
\begin{document}

\title{Performance Analysis of RIS-Assisted UAV Communication in NOMA Networks}

\author{
    \IEEEauthorblockN{Masoud Ghazikor\IEEEauthorrefmark{1}, Ly V. Nguyen\IEEEauthorrefmark{2}, Morteza Hashemi\IEEEauthorrefmark{1}}
    \IEEEauthorblockA{\IEEEauthorrefmark{1}Department of Electrical Engineering and Computer Science, University of Kansas,}
    \IEEEauthorblockA{\IEEEauthorrefmark{2}Elmore Family School of Electrical and Computer Engineering, Purdue University,}
}

\maketitle

\renewcommand\thefootnote{}\footnotetext{The material is based upon work supported by NSF grants 1955561, 2212565, 2323189, 2434113, and 2514415.}

\begin{abstract}
This paper investigates the performance of downlink non-orthogonal multiple access (NOMA) communication in unmanned aerial vehicle (UAV) networks enhanced by partitionable reconfigurable intelligent surfaces (RISs). We analyze three types of links between base station (BS) and UAVs: direct, RIS-only indirect, and composite links, under both Line-of-Sight (LoS) and Non-LoS (NLoS) propagation. The RIS-only indirect link and direct link are modeled using double Nakagami-$m$ and Nakagami-$m$ fading, respectively, while the composite link follows a combined fading channel model. Closed-form expressions for the cumulative distribution function (CDF) of the received signal-to-noise ratio (SNR) are derived for all links, enabling tractable outage probability analysis. Then, we formulate a \textit{fairness-efficiency bilevel optimization problem} to minimize the maximum outage probability among UAVs while minimizing the total number of required RIS reflecting elements. Accordingly, an \textit{RIS-assisted UAV Outage Minimization} (RUOM) algorithm is proposed, which fairly allocates the NOMA power coefficients while minimizing the total number of RIS reflecting elements required, subject to NOMA-defined constraints, RIS resource limitations, and maximum allowable outage threshold. Simulation results validate the analytical models and demonstrate that the proposed RUOM algorithm significantly improves fairness and efficiency in BS-UAV communication.
\end{abstract}


\section{Introduction}
Unmanned aerial vehicles (UAVs) have been experiencing rapid growth in the U.S. and globally, with Federal Aviation Administration (FAA) reports indicating over 1 million registered UAVs in the U.S. as of April 2025 \cite{FAA_UAV_2025}. Their expanding use in areas such as public safety, product delivery, and more highlights the need for safe and scalable integration into future airspace operations \cite{Badnava-2021-Spectrum}. As the number of UAVs and data traffic increases, ensuring reliable and efficient wireless communication becomes increasingly important, especially under limited spectrum resources. To address this, non-orthogonal multiple access (NOMA) has emerged as a key enabler in next-generation wireless networks, offering enhanced spectrum efficiency. In particular, power domain NOMA allows multiple users to share the same time and frequency resources by allocating different power levels \cite{Diao-2022-Enhancing}. This makes NOMA suitable for supporting the growing number of UAVs.

\begin{figure}[t]
\includegraphics[width=0.8\linewidth]{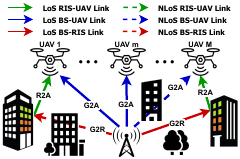}
\centering
\caption{\small System model consists of three types of links: direct (G2A), RIS-only indirect (G2R2A), and composite (G2A \& G2R2A), where a part of RIS elements (the green ones) is assigned to a specific UAV.}
\label{system_model}
\end{figure}

Despite the advantages of UAV-NOMA systems, they still face challenges such as severe path loss, signal blockage, and complex interference management. To overcome these, reconfigurable intelligent surfaces (RISs) have been proposed as a promising technique to enhance transmission performance. An RIS consists of a large number of passive elements that intelligently adjust their phase shifts to improve the wireless propagation environment. As a passive communication relay, the RIS consumes minimal energy, primarily for configuring its phase shifts, making it a highly energy-efficient technology for future wireless networks \cite{Liu-2023-Enabled}.

Recently, extensive research has been conducted on different aspects of RIS-assisted UAV-NOMA systems (see, for example, \cite{Tang-2025-Throughput, Feng-2023-Resource, Sobhi-2025-Efficient, Zhao-2022-RIS, Zhao-2025-Exploiting}). However, research gaps remain in developing RIS-assisted base station (BS)-UAV communication frameworks in NOMA systems that enhance fairness and efficiency among UAVs by jointly optimizing the power allocation and the number of assigned elements on a partitionable RIS. For instance, the work in \cite{Tang-2025-Throughput} investigates an RIS-assisted multi-UAV multi-user equipment (UE) NOMA communication system, maximizing system throughput by optimizing UAV trajectory, power allocation, and phase shift. In \cite{Feng-2023-Resource}, an RIS-assisted multi-UAV NOMA network is studied, minimizing the power consumption while satisfying minimum UE's data rate by optimizing UAV position, RIS coefficient, power allocation, beamforming, and decoding order. The study in \cite{Sobhi-2025-Efficient} analyzes the energy efficiency (EE) maximization by optimizing beamforming, phase shift, power allocation, and UAV placement in two scenarios: (1) UAV-mounted BS and (2) UAV-mounted RIS, each serving a NOMA cluster. In \cite{Zhao-2022-RIS}, an RIS-assisted uplink NOMA framework for UAVs and ground users in cellular networks is proposed, maximizing the network sum rate by optimizing UAV trajectory, RIS configuration, and power control.

Although there has been research on RIS-assisted UAV-NOMA systems, still remains a lack of thorough investigation into BS-UAV downlink communication using NOMA with partitionable RIS. To this end, we propose a comprehensive framework to analyze the RIS-assisted BS-UAV NOMA outage performance under Nagakami-$m$ and double Nakagami-$m$ fading channels. Then, we formulate a fairness-efficiency bilevel optimization problem to minimize the maximum outage probability to ensure fairness among UAVs, while minimizing the number of assigned RIS elements to ensure the outage probability remains under a predefined threshold, thereby improving efficiency. To solve this, we develop an iterative RIS-assisted UAV Outage Minimization (RUOM) algorithm, which fairly allocates the NOMA power coefficient, followed by efficient assignment of the required RIS elements. The main contributions of this paper are summarized as follows:
\begin{itemize}
    \item We present a comprehensive analytical framework to evaluate the outage probability of direct, RIS-only indirect, and composite links in partitionable RIS-assisted UAV-NOMA systems under Nakagami-$m$ and double Nakagami-$m$ fading channels. The proposed framework captures the dynamic characteristics of the UAV environment and their impact on outage performance.

    \item We propose the RUOM algorithm to solve the formulated fairness-efficiency bilevel optimization problem. In this iterative algorithm, the first part addresses the outer objective, which minimizes the maximum outage probability, ensuring fairness, while the second part handles the inner objective, which minimizes the total number of assigned reflecting elements, improving efficiency. 

    \item We demonstrate the effectiveness of the proposed RUOM algorithm in enhancing both fairness and efficiency among UAVs. Furthermore, we validate the analytical framework under various system settings.
\end{itemize}

\noindent The rest of this paper is organized as follows. In Section~\ref{systemmodel}, the system model is introduced, including the RIS-only indirect, direct, and composite channel models. Section~\ref{problemformulation} formulates a fairness-efficiency bilevel optimization problem and presents the RUOM algorithm. In Section \ref{numericalresults}, numerical results are provided, followed by the conclusion in Section \ref{conclusion}. 

\section{System Model} \label{systemmodel}
Assume a single-cell network where the centered-BS communicates in the downlink NOMA system by allocating power coefficients $\boldsymbol{\beta} = \{\beta_1, ..., \beta_M\}$ to $M$ UAVs, all operating in licensed bands specifically allocated for UAV operations \cite{FCC_Spectrum_2025}. As illustrated in Fig. \ref{system_model}, BS can establish three types of links: 1. direct: Ground-to-Air (G2A), 2. RIS-only indirect: Ground-to-RIS-to-Air (G2R2A), and 3. composite: G2A \& G2R2A, which can be either a Line-of-Sight (LoS) link or a Non-LoS (NLoS) link. In the RIS-assisted links, a partitionable RIS $k$ is considered, where part of it can be assigned to UAV $m$, with the number of assigned elements is denoted by $N_m^k$.
\begin{figure*}[t]
\small
\begin{equation} \label{compositecdf}
F_{\gamma_{c}}(\gamma) =
\begin{cases}
\displaystyle
\Big(1 - Q(\sqrt{a})^{-1} \Big) F_{\gamma_d}(\gamma)
+ \frac{Q(\sqrt{a})^{-1}}{\Gamma(m_3)} \left( \frac{m_3}{\Omega_3 |\hat{g}^d|^2} \right)^{m_3}
\psi\left(0, \sqrt{\frac{\gamma}{\Bar{\gamma}_c}}, \boldsymbol{c} \right),
& \mathbb{E} [\Tilde{\boldsymbol{g}}^{r}] > \sqrt{\frac{\gamma}{\Bar{\gamma}_r}} \\[12pt]

\begin{aligned}
\displaystyle
&
\Big(1 - Q(\sqrt{a})^{-1} \Big) F_{\gamma_d}(\gamma)
+ \frac{Q(\sqrt{a})^{-1}}{\Gamma(m_3)} \, \gamma_{inc}\left(m_3, \frac{m_3 (\sqrt{\frac{\gamma}{\Bar{\gamma}_r}} - \mathbb{E} [\Tilde{\boldsymbol{g}}^{r}])^2}{\Omega_3 |\hat{g}^d|^2} \right) + \frac{Q(\sqrt{a})^{-1}}{\Gamma(m_3)} \\ & \times \left( \frac{m_3}{\Omega_3 |\hat{g}^d|^2} \right)^{m_3} \left[ \psi\left(\sqrt{\frac{\gamma}{\Bar{\gamma}_r}} - \mathbb{E} [\Tilde{\boldsymbol{g}}^{r}], \sqrt{\frac{\gamma}{\Bar{\gamma}_c}}, \boldsymbol{c} \right)
- \psi\left(0, \sqrt{\frac{\gamma}{\Bar{\gamma}_r}} - \mathbb{E} [\Tilde{\boldsymbol{g}}^{r}], \boldsymbol{c} \right) \right],
\end{aligned}
& \mathbb{E} [\Tilde{\boldsymbol{g}}^{r}] \leq \sqrt{\frac{\gamma}{\Bar{\gamma}_r}}
\end{cases}
\end{equation}
\end{figure*}

\begin{figure*}[t]
\small
\begin{equation*}
\psi(p_1, p_2, \boldsymbol{c}) =
\begin{cases}
\displaystyle\sum_{i=0}^{2m_3 - 1} \frac{\rho_i}{2c_1^{\frac{i+1}{2}}}
\left[ \Gamma_{inc}\left( {\frac{i+1}{2}}, c_1(p_1 - \frac{c_2}{c_1})^2 \right) - \Gamma_{inc}\left( {\frac{i+1}{2}}, c_1(p_2 - \frac{c_2}{c_1})^2 \right) \right],
& p_1 > \frac{c_2}{c_1} \text{ or } i \text{ odd} \\[12pt]

\displaystyle\sum_{i=0}^{2m_3 - 1} \frac{\rho_i}{2c_1^{\frac{i+1}{2}}}
\left[ \Gamma_{inc}\left( {\frac{i+1}{2}}, c_1(p_2 - \frac{c_2}{c_1})^2 \right) - \Gamma_{inc}\left( {\frac{i+1}{2}}, c_1(p_1 - \frac{c_2}{c_1})^2 \right) \right],
& p_2 < \frac{c_2}{c_1} \text{ and } i \text{ even} \\[12pt]

\displaystyle\sum_{i=0}^{2m_3 - 1} \frac{\rho_i}{2c_1^{\frac{i+1}{2}}}
\left[ \gamma_{inc}\left( {\frac{i+1}{2}}, c_1(p_1 - \frac{c_2}{c_1})^2 \right) + \gamma_{inc}\left( {\frac{i+1}{2}}, c_1(p_2 - \frac{c_2}{c_1})^2 \right) \right],
& p_1 < \frac{c_2}{c_1} < p_2 \text{ and } i \text{ even}
\end{cases}
\end{equation*}
\begin{equation*}
    \rho_i = {2m_3-1 \choose i} \exp{ \Big(\frac{c_2^2}{c_1} - c_3 \Big)} \Big(\frac{c_2}{c_1} \Big)^{2m_3 - i - 1},
    \qquad
    \boldsymbol{c} = \Big(\frac{m_3}{\Omega_3 |\hat{g}^d|^2} + \frac{1}{|\hat{g}^r|^2 \sigma^2_{\Tilde{\boldsymbol{g}}^{r}}}, \frac{\sqrt{\frac{\gamma}{\Bar{\gamma}_r}} - \mathbb{E} [\Tilde{\boldsymbol{g}}^{r}]}{|\hat{g}^r|^2 \sigma^2_{\Tilde{\boldsymbol{g}}^{r}}}, \frac{(\sqrt{\frac{\gamma}{\Bar{\gamma}_r}} - \mathbb{E} [\Tilde{\boldsymbol{g}}^{r}])^2}{|\hat{g}^r|^2 \sigma^2_{\Tilde{\boldsymbol{g}}^{r}}} \Big),
\end{equation*}
\end{figure*}

\noindent
\textbf{RIS-only Channel Model.} Consider $P_r = P_t |g^{r}|^2$ with the received power $P_r$ and transmit power $P_t$, in which, the channel gain is defined as $g^{r} = \sum_{i=1}^{N} g^{g}_{i} e^{\mathrm{j}\Phi_i} g^{a}_{i}$, where $\Phi_i$ denotes the RIS phase shift by the $i$-th reflecting element \cite{Yang-2020-Performance}. Also, $g^{a}_{i} = \Tilde{g}_i^{a} e^{\mathrm{j}\phi_i} \hat{g}^{a}$ and $g^{g}_{i} = \Tilde{g}_i^{g} e^{\mathrm{j}\varphi_i} \hat{g}^{g}$ are RIS-to-Air (R2A) and Ground-to-RIS (G2R) channel gains, respectively. Furthermore, $\Tilde{g}_i$ and $\hat{g} = \sqrt{d^{-\alpha(d)}}$ are the fading coefficients modeled by Nakagami-$m$ fading and the square root of path loss model, equal for $N$ reflecting elements \cite{Zhang-2022-Joint}. Moreover, the path loss exponent (PLE) is expressed as $\alpha(d) = \alpha_{L}\mathbb{P}_{L}(d) + \alpha_{N}(1 - \mathbb{P}_{L}(d))$, where $\alpha_{L}$ and $\alpha_{N}$ denote the PLE for LoS and NLoS channels, respectively. Also, LoS probability $\mathbb{P}_{L}(d)$ is defined as \cite{Kim-2019-Impact}:
\begin{align*}
\mathbb{P}_{L}(d) = 
\begin{cases}
    \big(1-\exp{(-\frac{z_{1}^2}{2\zeta^2})}\big)^{d\sqrt{v\mu}} & \text{$z_{1} = z_{2}$}\\
    \big(1-\frac{\sqrt{2\pi}\zeta}{d^{V}}|Q(\frac{z_{1}}{\zeta})-Q(\frac{z_{2}}{\zeta})|\big)^{d^{H}\sqrt{v\mu}} & \text{$z_{1} \neq z_{2}$},
\end{cases}
\end{align*}
where $\zeta$, $v$, and $\mu$ are environmental parameters, also $Q(.)$, $d^H$, and $d^V$ are the $Q$-function, horizontal and vertical distances, respectively. The optimal RIS phase shift is $\Phi_i = -(\phi_i + \varphi_i)$ to achieve the maximum SNR. Thus, the RIS channel gain can be rewritten as $g^{r} = \hat{g}^{g} \hat{g}^{a} \sum_{i = 1}^{N} \Tilde{g}_i^{r} = \hat{g}^{r} \Tilde{\boldsymbol{g}}^{r}$, where $\Tilde{g}_i^{r} = \Tilde{g}_i^{g} \Tilde{g}_i^{a}$ represents the G2R2A fading coefficient associated with the $i$-th reflecting element. Accordingly, $\Tilde{g}_{i}^{r}$ is modeled by double Nakagami-$m$ fading, and the PDF of $\Tilde{g}_{i}^{r}$ is given by:
\begin{equation}
f_{\Tilde{g}_{i}^{r}}(x) = \frac{4x^{m_1+m_2-1} K_{m_1-m_2} \Big(2x \sqrt{\frac{m_1m_2}{\Omega_1 \Omega_2}} \Big) }{\Gamma(m_1) \Gamma(m_2) (\frac{\Omega_1 \Omega_2}{m_1 m_2})^{\frac{m_1+m_2}{2}}},
\end{equation}
where $\Gamma(.)$, $\Omega$, and $K_v(.)$ are the gamma function, average fading power, and $v^{th}$-order modified Bessel function of the second kind, respectively \cite{Tegos-2022-Distribution}. Also, $m$ denotes the shape parameter, which is approximated as $m \approx \frac{(\exp(2.708\mathbb{P}_{L}(d)^2) + 1)^2}{2 \exp(2.708\mathbb{P}_{L}(d)^2) + 1}$ using the Rician $K$-factor \cite{Channel_Ghazikor_2024}. Moreover, the $n^{th}$ order moment of $\Tilde{g}_{i}^{r}$ is expressed as
\begin{equation}
    \mathbb{E} [\Tilde{g}_{i}^{r^n}] = \prod_{j=1}^2 \frac{\Gamma(m_j + \frac{n}{2})}{\Gamma(m_j)} \Big(\frac{\Omega_j}{m_j} \Big)^{\frac{n}{2}}.
\end{equation}
Applying the first term of the Laguerre series, the PDF of $\Tilde{\boldsymbol{g}}^{r}$ over $N$ reflecting elements can be approximated as \cite{Abualhayja-2024-Exploiting, Dorra-2011-SIR}:
\begin{equation}
f_{\Tilde{\boldsymbol{g}}^{r}}(x) = \frac{x^{a-1} \exp{(-\frac{x}{b})}}{b^a\Gamma(a)}.
\end{equation}
Here, $a = \frac{\mathbb{E} [\Tilde{\boldsymbol{g}}^{r}]^2}{\sigma^2_{\Tilde{\boldsymbol{g}}^{r}}} = \frac{N \mathbb{E} [\Tilde{g}_{i}^{r}]^2}{\mathbb{E} [\Tilde{g}_{i}^{r^2}] - \mathbb{E} [\Tilde{g}_{i}^{r}]^2}$ and $b = \frac{\sigma^2_{\Tilde{\boldsymbol{g}}^{r}}}{\mathbb{E} [\Tilde{\boldsymbol{g}}^{r}]} = \frac{\mathbb{E} [\Tilde{g}_{i}^{r^2}] - \mathbb{E} [\Tilde{g}_{i}^{r}]^2}{\mathbb{E} [\Tilde{g}_{i}^{r}]}$ where $\Tilde{g}_{i}^{r}$ is independent for $N$ reflecting elements. Thus, the PDF and CDF of the maximum instantaneous SNR $\gamma_{r} = \frac{P_t |g^{r}|^2}{P_N} = \frac{P_t |\hat{g}^{r}|^2}{P_N} |\Tilde{\boldsymbol{g}}^{r}|^2 = \Bar{\gamma}_{r} |\Tilde{\boldsymbol{g}}^{r}|^2$ are written as \cite{Yang-2022-Performance}:
\begin{equation}
f_{\gamma_{r}}(\gamma) = \frac{(\frac{\gamma}{\Bar{\gamma}_{r}})^{\frac{a}{2}-1} \exp{(-\sqrt{\frac{\gamma}{\Bar{\gamma}_{r} b^2}})}}{2\Bar{\gamma}_{r}b^a\Gamma(a)},
\end{equation}
\begin{equation} \label{riscdf}
F_{\gamma_{r}}(\gamma) = \frac{\gamma_{inc}(a, \sqrt{\frac{\gamma}{\Bar{\gamma}_{r} b^2}})}{\Gamma(a)},
\end{equation}
where $\gamma_{inc}(., .)$ denotes the lower incomplete gamma function and $P_N = \kappa TB$ is the thermal noise power where $\kappa$, $T$, and $B$ are Boltzmann’s constant, temperature, and bandwidth.

\noindent
\textbf{Direct Channel Model.} Similar to the RIS channel model, the direct G2A channel gain $g^{d}$ can be expressed as $g^{d} = \hat{g}^{d} e^{\mathrm{j}\theta} \Tilde{g}^{d}$, where the fading coefficient $\Tilde{g}^{d}$ can be modeled by Nakagami-$m$ fading that has the following PDF \cite{Abualhayja-2024-Exploiting}:
\begin{equation}
f_{\Tilde{g}^{d}}(x) = \frac{2m_3^{m_3} x^{2m_3-1} \exp{(-\frac{m_3x^2}{\Omega_3})}}{\Omega_3^{m_3} \Gamma(m_3)}.
\end{equation}
Accordingly, the PDF and CDF of the instantaneous SNR $\gamma_{d} = \frac{P_t |\hat{g}^{d}|^2}{P_N}|\Tilde{g}^{d}|^2 = \Bar{\gamma}_{d} |\Tilde{g}^{d}|^2$ are given by \cite{Alqahtani-2021-Performance}:
\begin{equation}
f_{\gamma_{d}}(\gamma) = \frac{m_3^{m_3} (\frac{\gamma}{\Bar{\gamma}_{d}})^{m_3-1} \exp{(-\frac{m_3\gamma}{\Omega_3 \Bar{\gamma}_{d}})}}{\Bar{\gamma}_{d} \Omega_3^{m_3} \Gamma(m_3)}
\end{equation}
\begin{equation}
F_{\gamma_{d}}(\gamma) = \frac{\gamma_{inc}(m_3, \frac{m_3\gamma}{\Omega_3 \Bar{\gamma}_{d}})}{\Gamma(m_3)}.
\end{equation}
Notably, the Rayleigh fading channel for ground-level communication can also be obtained when $m_3 = 1$.

\noindent
\textbf{Composite Channel Model.} The composite link combines RIS-only indirect and direct links. Given that the optimal RIS phase shifts are $\Phi_i = -(\phi_i + \varphi_i - \theta)$, the SNR for the composite channel is expressed as $\gamma_c = \frac{P_t |g^c|^2}{P_N} = \Bar{\gamma}_c (|g^{r}| + |g^{d}|)^2$, where the composite channel gain $g^c$ consists of the RIS-only indirect channel gain $g^r$ and direct channel gain $g^d$. Thus, the CDF of the maximum instantenous SNR $\gamma_{c}$ is defined as:
\begin{equation}
    F_{\gamma_c}(\gamma) = \int^{\sqrt{\frac{\gamma}{\Bar{\gamma}_c}}}_{0} F_{\gamma_r} \Big(\sqrt{\frac{\gamma}{\Bar{\gamma}_c}} - x \Big) f_{\gamma_d}(x) dx.
\end{equation}
For the tractability of analysis, the CDF of ${\gamma_r}$ in Eq. \eqref{riscdf} is approximated using the $Q$-function as:
\begin{equation}
    F_{\gamma_r}(\gamma) \approx 1 - \frac{Q \Big(\frac{\sqrt{\frac{\gamma}{\Bar{\gamma}_r}} - \mathbb{E} [\Tilde{\boldsymbol{g}}^{r}]}{ \sqrt{\sigma^2_{\Tilde{\boldsymbol{g}}^{r}}}} \Big)}{Q(\sqrt{a})}.
\end{equation}

This CDF is also proved by the central limit theorem (CLT) in \cite{Analysis_Ni_2023}. Finally, the CDF of $\gamma_c$ can be rewritten as:
\begin{align*}
    F_{\gamma_c}(\gamma) = F_{\gamma_d}(\gamma) - \int_{0}^{\sqrt{\frac{\gamma}{\Bar{\gamma}_c}}} \frac{Q(\frac{\sqrt{\frac{\gamma}{\Bar{\gamma}_c}} - x - \hat{g}^r \mathbb{E} [\Tilde{\boldsymbol{g}}^{r}]}{\hat{g}^r \sqrt{\sigma^2_{\Tilde{\boldsymbol{g}}^{r}}}})}{Q(\sqrt{a})} f_{\gamma_d}(x) dx,
\end{align*}
where the integral solution can be found in \cite{Analysis_Ni_2023} by dividing it into two cases when $\mathbb{E} [\Tilde{\boldsymbol{g}}^{r}] > \sqrt{\frac{\gamma}{\Bar{\gamma}_r}}$ or $\mathbb{E} [\Tilde{\boldsymbol{g}}^{r}] \le \sqrt{\frac{\gamma}{\Bar{\gamma}_r}}$. Finally, the CDF of $\gamma_c$ can be expressed as Eq. \eqref{compositecdf}.

\noindent
\textbf{NOMA Outage Probability.} Consider $M$ UAVs in the downlink NOMA system where UAVs are ordered according to the channel gains such as $|g_1|^2 \le ... \le |g_m|^2 \le ... \le |g_M|^2$. Accordingly, the power coefficients for $M$ UAVs are allocated as $\beta_1 > ... > \beta_m > ... > \beta_M$ such as $\sum_{i=1}^{M} \beta_i = 1$. Assuming perfect successive interference cancellation (SIC), UAV $m$ can subtract the first $m-1$ UAV signals from the transmitted superimposed signal and treat the other UAV signals $m<i\le M$ as noise \cite{Chen-2024-Optimizing}. Therefore, the achievable data rate on UAV $m$ is defined as: 
\begin{equation}
    R_m = \log_2 \Big(1+\frac{\gamma_m \beta_m}{\gamma_m \sum_{i=m+1}^{M}\beta_i + 1} \Big),
\end{equation}
where $\gamma_m = \frac{\eta_m}{\beta_m} = \frac{P_t |g_m|^2}{P_N}$ denotes the SNR of the $m$-th ordered UAV. Note that $R_m$ must satisfy the condition $R_{m,j} \geq R^{tg}_j$, where $R^{tg}_j$ is the target data rate for UAV $j$, which is measured in terms of bits per channel use (bpc), corresponding to its quality of service (QoS) requirement. Accordingly, $R_{m,j}$ represents UAV $m$ data rate to decode the UAV $j$'s signal, $j \leq m$, which can be expressed as:
\begin{equation}
    R_{m,j} = \log_2 \Big(1+\frac{\gamma_m \beta_j}{\gamma_m \sum_{i=j+1}^{M}\beta_i + 1} \Big).
\end{equation}
By employing the order statistics approach to model the differences for channel gains in NOMA systems, the PDF and CDF of $\gamma_m$, $1 \le m \le M$, is defined as \cite{Alqahtani-2021-Performance, David-2003-Order}:
\begin{align*}
    f_{\gamma_m}(\gamma) = m {M\choose m} f_{\gamma}(\gamma) (F_{\gamma}(\gamma))^{m-1} (1-F_{\gamma}(\gamma))^{M-m},
\end{align*}
\begin{align*}
    F_{\gamma_m}(\gamma) = m {M\choose m} \sum^{M-m}_{n=0} {M-m \choose n} \frac{(-1)^n (F_{\gamma}(\gamma))^{m+n}}{m+n}.
\end{align*} 
Thus, the NOMA-based outage probability $\mathbb{P}_m^{out}(\boldsymbol{\beta}, N^k_m)$ for UAV $m$ can be expressed as follows:
\begin{equation} \label{noma-gra-out}
    \mathbb{P}_m^{out}(\boldsymbol{\beta}, N^k_m) = 1 - \mathbb{P} \Big( (R_{m,1} > R_1^{tg}) \boldsymbol{\cap} ... \boldsymbol{\cap} (R_{m,m} > R_m^{tg}) \Big),
\end{equation}
where $\mathbb{P}(R_{m,j} > R_j^{tg})$ for $1 \le j \le m$ can be simplified as:
\begin{align*}
    \begin{aligned}
        & \mathbb{P}(R_{m,j} > R_j^{tg}) = \mathbb{P} \Big(\log_2 \big( 1+\frac{\gamma_m \beta_j}{\gamma_m \sum_{i=j+1}^{M} \beta_i + 1} \big) > R_j^{tg} \Big) \\ & 
        = \mathbb{P}(\gamma_m > \frac{2^{R_j^{tg}} - 1}{\beta_j - (2^{R_j^{tg}} - 1) \sum_{i=j+1}^M \beta_i})
        = \mathbb{P}(\gamma_m > \gamma_j^{lb}),
    \end{aligned}
\end{align*}
where Eq. \eqref{noma-gra-out} is constrained by $(2^{R_j^{tg}} - 1) \sum_{i=j+1}^M \beta_i < \beta_j$ and $\gamma^{mlb}_m = \max\{\gamma_1^{lb}, \dots, \gamma_m^{lb}\}$ for $m<M$, while 
$\gamma^{mlb}_M = \gamma_M^{lb}$. Hence, Eq. \eqref{noma-gra-out} can be rewritten as:
\begin{align}
\begin{aligned}
    & \mathbb{P}_m^{out}(\boldsymbol{\beta}, N^k_m) = 1 - \mathbb{P}(\gamma_m > \gamma^{mlb}_m) = F_{\gamma_m}(\gamma^{mlb}_m) \\ &
    = m {M\choose m} \sum^{M-m}_{n=0} {M-m \choose n} \frac{(-1)^n (F_{\gamma}(\gamma^{mlb}_m))^{m+n}}{m+n}.
\end{aligned}
\end{align}

\section{Problem Formulation and Proposed Solution} \label{problemformulation}
In this section, we formulate a fairness-efficiency bilevel optimization problem, then we propose the RUOM algorithm (Algorithm \ref{alg:ruom}), followed by a progressive grid search (PGS) sub-algorithm (Algorithm \ref{alg:pgs}) to solve the formulated problem.

\begin{algorithm}[t]
\caption{\small RIS-assisted UAV Outage Minimization (RUOM).}
\small
\label{alg:ruom}
\begin{algorithmic}[1]
\Function{RUOM}{$\lambda$, $\delta$, $\epsilon^{in}$, $\epsilon^{ac}$}
    \State \textbf{Initialize} $N_m^k \gets 0$, $\forall m,\; k$, $\boldsymbol{\beta}^0 \gets \text{None}$
    \For{$t = 1$ to $\mathcal{T}$}
        \State $\epsilon^{sr} \gets \epsilon^{in}$, $\boldsymbol{\beta}^t \gets \text{None}$ \Comment{Fairness}
        \While{$\epsilon^{sr} > \epsilon^{ac}$}
            \State $\mathcal{B} \gets \Call{PGS}{\boldsymbol{\beta}^t, \epsilon^{sr}}$
            \State $\boldsymbol{\beta}^t \gets \arg\min_{\boldsymbol{\beta} \in \mathcal{B}} \left( \max_{m} \mathbb{P}_m^{out}(\boldsymbol{\beta}, N_m^k) \right)$ \textbf{in parallel}
            \State $\epsilon^{sr} \gets \lambda \cdot \epsilon^{sr}$
        \EndWhile
        \For{$m = 1$ to $M$} \Comment{Efficiency}
            \While{$\mathbb{P}_m^{out}(\boldsymbol{\beta}^t, N_m^k) < \delta$ \textbf{\&} $N_m^k \ge 1$}
                \State $N_m^k \gets N_m^k - 1$
            \EndWhile
            \While{$\mathbb{P}_m^{out}(\boldsymbol{\beta}^t, N_m^k) \ge \delta$ \textbf{\&} $\sum\limits^M_{m=1} N_m^k \leq \mathcal{N}^k$}
                \State $N_m^k \gets N_m^k + 1$
            \EndWhile
        \EndFor
        \If{$\|\boldsymbol{\beta}^{t} - \boldsymbol{\beta}^{t-1}\| < \epsilon$}
            \State $\boldsymbol{\beta}^{\star} \gets \boldsymbol{\beta}^t$
            \State \textbf{break}
        \EndIf
        \State $\boldsymbol{\beta}^{t-1} \gets \boldsymbol{\beta}^t$
    \EndFor
    \State \textbf{return} $\boldsymbol{\beta}^\star, N_m^k$
\EndFunction
\end{algorithmic}
\end{algorithm}

\noindent
\textbf{Problem Formulation.} As mentioned earlier, our objective is to minimize the maximum outage probability $\mathbb{P}_m^{out}(\boldsymbol{\beta}, N_m^k)$ for UAV $m$ by fairly adjusting the power coefficient vector $\boldsymbol{\beta}$ while efficiently assigning the minimum number of RIS reflecting elements $N^k_m$ to maintain the outage probability $\mathbb{P}_m^{out}(\boldsymbol{\beta}, N_m^k)$ under a threshold $\delta$. This can be formulated as a fairness-efficiency bilevel optimization problem as follows:
\begin{subequations}
\begin{align}
\min_{\boldsymbol{\beta}} \ & \max_{m} \mathbb{P}_m^{out}(\boldsymbol{\beta}, {N_m^k}^{*}) \label{eq:beta_outer_obj} \\
\text{s.t.} & 
\sum_{m=1}^{M} \beta_m = 1, \label{eq:beta_cons1} \\ 
& (2^{R_j^{tg}} - 1) \sum_{i=j+1}^{M} \beta_i < \beta_j, \quad 1 \le j \le m, \label{eq:beta_cons2} \\
& \{{N_m^k}^{*}\} \in \arg \min_{\{N_m^k\}} \quad \sum_{m=1}^{M} \sum_{k=1}^{K} N_m^k \label{eq:ris_inner_obj} \\
& \text{s.t. } \mathbb{P}_m^{out}(\boldsymbol{\beta}, N_m^k) < \delta, \quad \forall m, \label{eq:ris_cons1} \\
& \quad \sum_{m=1}^{M} N_m^k \leq \mathcal{N}^k, \quad \forall k, \label{eq:ris_cons2}
\end{align}
\end{subequations}
where the outer objective \eqref{eq:beta_outer_obj} guarantees fairness with the constraints \eqref{eq:beta_cons1} and \eqref{eq:beta_cons2} on the power coefficients $\boldsymbol{\beta}$, which introduced in NOMA definition. Also, the inner objective \eqref{eq:ris_inner_obj} ensures efficiency with the constraints \eqref{eq:ris_cons1} and \eqref{eq:ris_cons2}, which maintain the outage probability under a threshold $\delta$ and limit the number of assigned reflecting elements on RIS $k$, respectively. 

\begin{figure*}[t]
    \centering
    \begin{minipage}{.33\linewidth}
        \includegraphics[width=6cm]{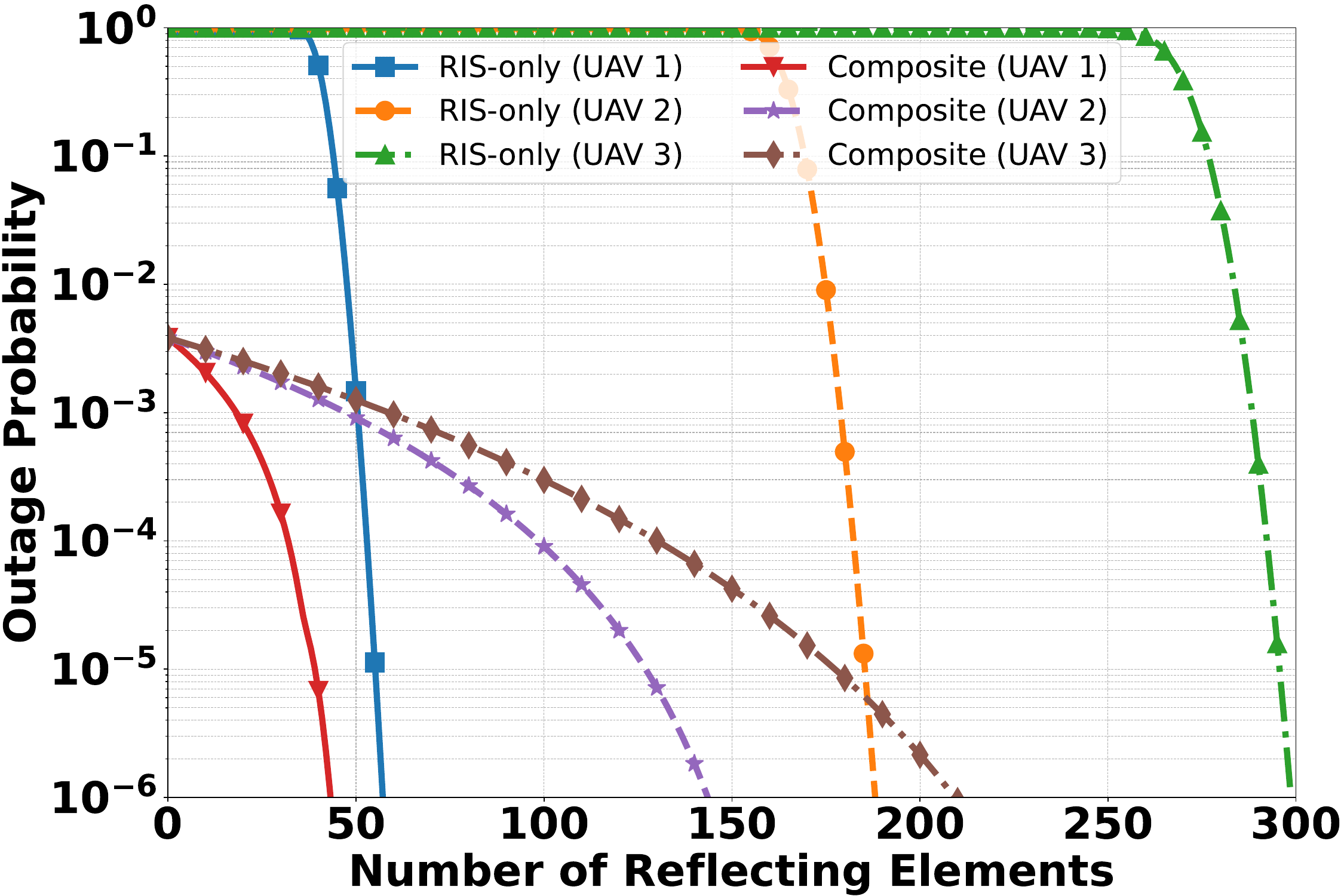}
        \caption{$\mathbb{P}_m^{out}(\boldsymbol{\beta}, N_m^{k^\star})$ vs. various links.}
        \label{fig:out_link}
    \end{minipage}%
    \begin{minipage}{.33\linewidth}
        \includegraphics[width=6cm]{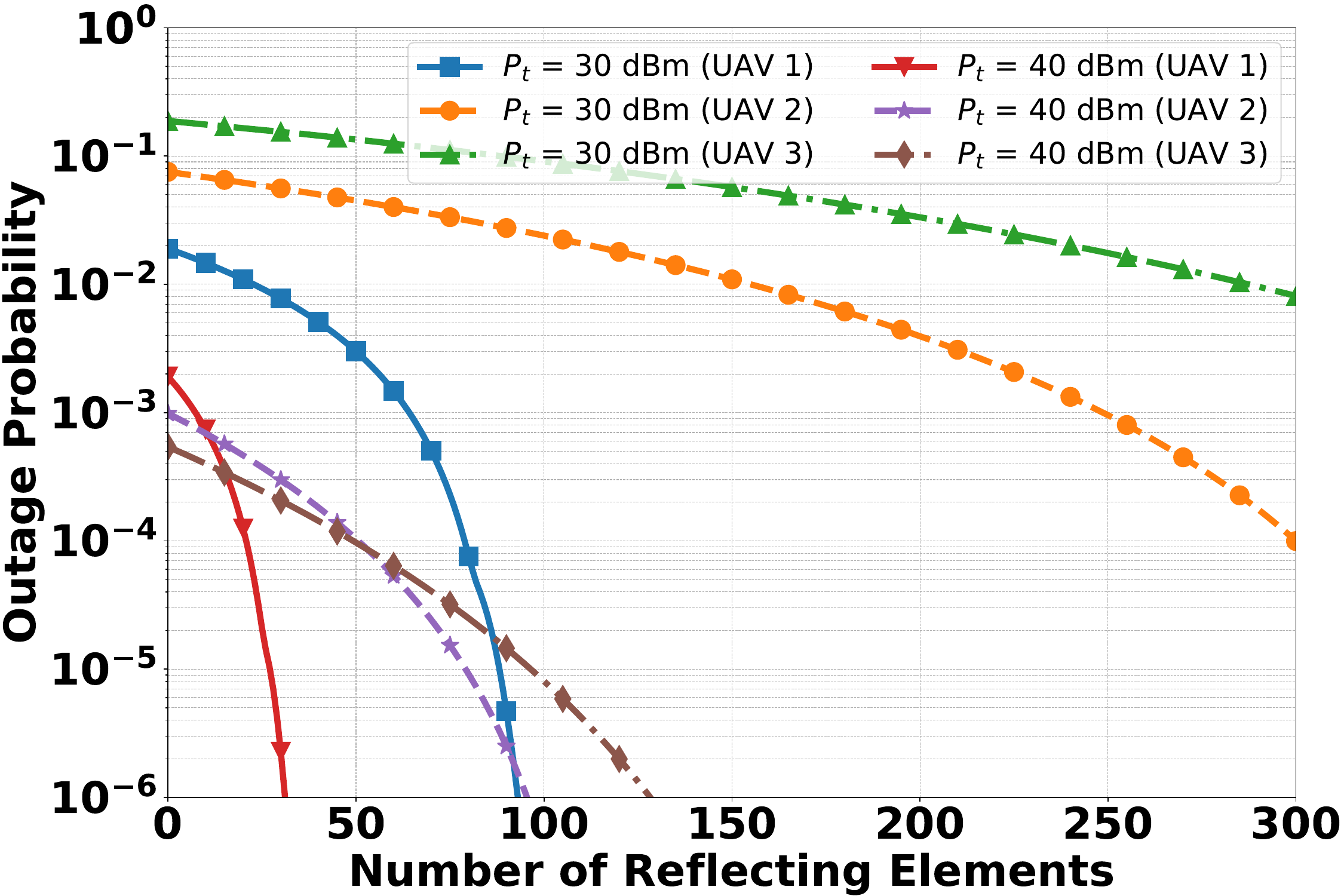}
        \caption{$\mathbb{P}_m^{out}(\boldsymbol{\beta}, N_m^{k^\star})$ vs. $P_t$.}
        \label{fig:out_power}
    \end{minipage}%
    \begin{minipage}{.33\linewidth}
        \includegraphics[width=6cm]{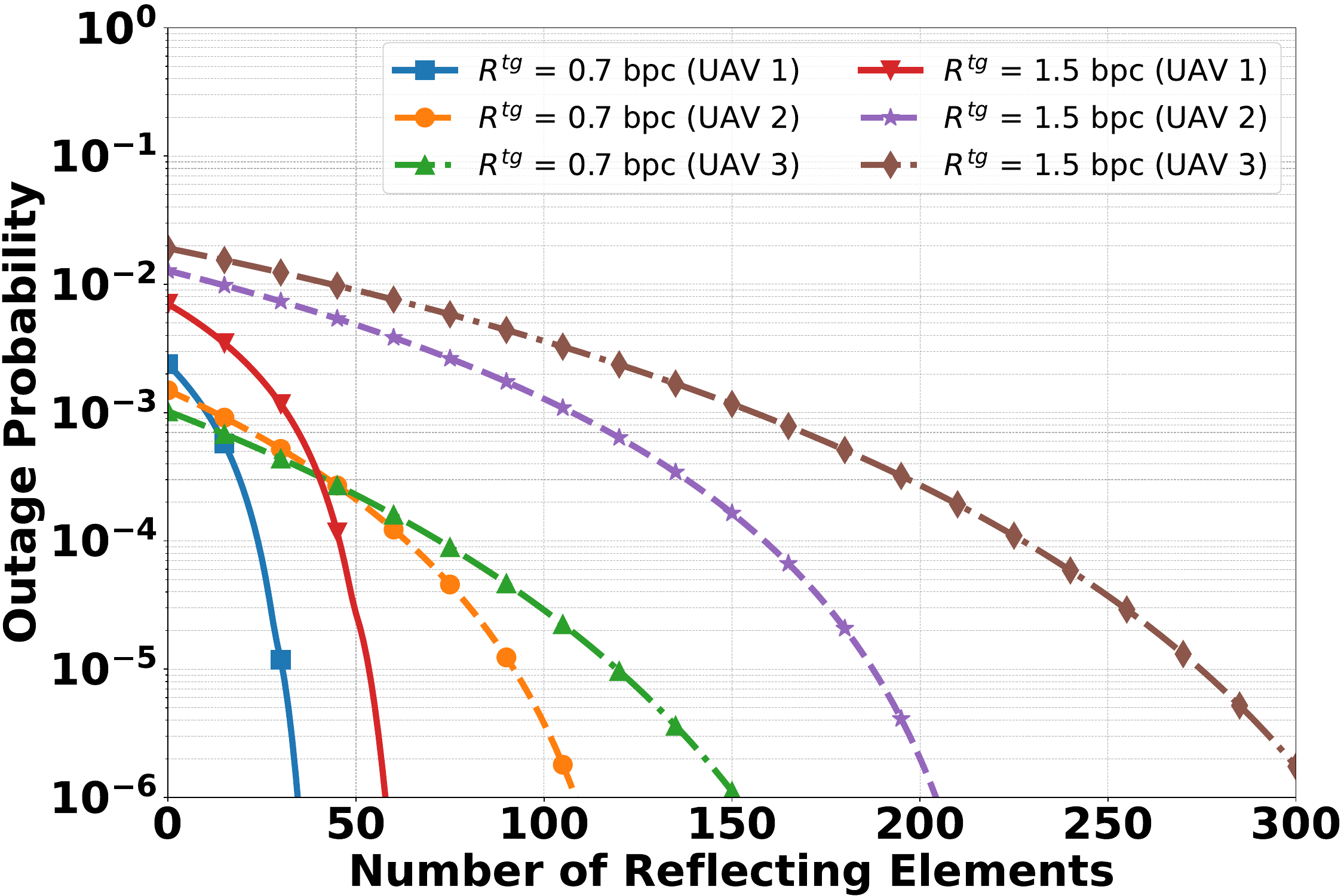}
        \caption{$\mathbb{P}_m^{out}(\boldsymbol{\beta}, N_m^{k^\star})$ vs. $R^{tg}$.}
        \label{fig:out_rate}
    \end{minipage}%
\end{figure*}

\noindent
\textbf{Proposed RUOM Algorithm.} To solve our formulated fairness-efficiency bilevel optimization problem, we propose the iterative RUOM algorithm, including fairness and efficiency parts, as presented in Algorithm \ref{alg:ruom}:
\begin{itemize}
    \item \textbf{Fairness (lines 4-9):} In this part, the outer objective is to minimize the maximum outage probability $\mathbb{P}_m^{out}(\boldsymbol{\beta}^t, N_m^{k})$ by finding the best power coefficient vector $\boldsymbol{\beta}^t$ at iteration $t$ from valid combinations of $\boldsymbol{\beta}$ in a set $\mathcal{B}$, generated by the PGS sub-algorithm. Within the while loop (line 5), the best $\boldsymbol{\beta}^t$ is iteratively refined by selecting the candidate that achieves the lowest maximum outage probability $\mathbb{P}_m^{out}(\boldsymbol{\beta}^t, N_m^{k})$ across all $M$ UAVs. The accuracy of this refinement depends on the search resolution $\epsilon^{sr}$ scaled by a factor $\lambda$ until it exceeds the accuracy threshold $\epsilon^{ac}$. 
    
    \textbf{PGS Sub-Algorithm:} The PGS sub-algorithm generates a discrete set $\mathcal{B}$ of candidate power coefficient vectors $\boldsymbol{\beta}$ with search resolution $\epsilon^{sr}$. Without prior $\boldsymbol{\beta}$ solution ($\boldsymbol{\beta} = $ None), it creates a global search space over the full grid $\mathcal{G}^M$; otherwise, it builds a local search space around the current vector $\boldsymbol{\beta}$ by restricting each $\beta_m$ to a bounded interval intersected with the global grid. Finally, each candidate vector $\boldsymbol{\beta}$ is verified to satisfy the constraints \eqref{eq:beta_cons1} and \eqref{eq:beta_cons2}, and the feasible set $\mathcal{B}$ is returned.
    
    \item \textbf{Efficiency (lines 10–17):} In this part, the inner objective is to minimize the total number of assigned reflecting elements while ensuring that the outage probability $\mathbb{P}_m^{out}(\boldsymbol{\beta}^t, N_m^{k})$ for each UAV $m$ remains below a threshold $\delta$. For all $M$ UAVs, it first attempts to reduce the assigned RIS reflecting elements $N_m^k$ as long as the constraint \eqref{eq:ris_cons1} is satisfied. If the outage exceeds $\delta$, it gradually increases $N_m^k$ until the constraint \eqref{eq:ris_cons1} is met, subject to the maximum number of reflecting elements $\mathcal{N}^k$ in constraint \eqref{eq:ris_cons2}. This process guarantees efficient RIS resource usage while ensuring reliability for all $M$ UAVs.
\end{itemize}
At the end of each iteration $t$, if the power coefficient vector $\boldsymbol{\beta}^t$ changes by less than $\epsilon$ from the previous iteration, i.e., $\|\boldsymbol{\beta}^{t} - \boldsymbol{\beta}^{t-1}\| < \epsilon$, it indicates that the power coefficient vector $\boldsymbol{\beta}^{t}$ has converged and no further adjustment to the assigned RIS reflecting elements is necessary; therefore, the algorithm stops and returns the optimal $\boldsymbol{\beta}^{\star}$ and the optimal number of RIS elements $N^k_m$.

\begin{algorithm}[t]
\caption{Progressive Grid Search (PGS)}
\small
\label{alg:pgs}
\begin{algorithmic}[1]
\Function{PGS}{$\boldsymbol{\beta}, \epsilon^{sr}$}
    \State $\mathcal{B} \gets \emptyset$
    \If{$\boldsymbol{\beta} = \text{None}$} \Comment{Global Space}
        \State $\mathcal{C} \gets \left\{ \boldsymbol{\beta} \in \mathcal{G}^M \right\}$
        \Comment{$\mathcal{G} = \{0, \epsilon^{sr}, 2\epsilon^{sr}, \dots, 1\}$}
    \Else \Comment{Local Space}
        \State $a_m \gets \max(0,\; \beta_m - \epsilon^{sr}), b_m \gets \min(1,\; \beta_m + \epsilon^{sr})$
        \State $\mathcal{C} \gets \prod\limits_{m=1}^M \left( [a_m,\; b_m] \cap \mathcal{G} \right)$
    \EndIf
    \ForAll{$\boldsymbol{\beta} \in \mathcal{C}$}
        \If{$\sum\limits^M_{m=1} \beta_m = 1$ \textbf{\&} $(2^{R_j^{tg}} - 1) \sum\limits_{i=j+1}^M \beta_i < \beta_j$}
            \State $\mathcal{B} \gets \mathcal{B} \cup \{ \boldsymbol{\beta} \}$
        \EndIf
    \EndFor
    \State \Return $\mathcal{B}$
\EndFunction
\end{algorithmic}
\end{algorithm}

\section{Numerical Results} \label{numericalresults}

In our simulation, 3 UAVs and 3 RISs are placed according to the Poisson distribution with UAVs' altitude in the limited range of 80 m to 120 m, and the centered-BS communicates with each UAV $m$ both directly and through the best RIS $k^{\star}$.
Key simulation parameters are summarized in Table \ref{tab:sim_parameters}.

\begin{table} [t]
    \centering
    \small
    \caption{Key Simulation Parameters}
    \resizebox{\columnwidth}{!}{%
    \label{tab:sim_parameters}
    \begin{tabular}{lc}
        \toprule
        Definition & Notation \& Value \\
        \midrule
        Cell Radius & $r = 2$ km \\
        Environment Parameters & $\zeta = 20$, $v = 3 \times 10^{-4}$, $\mu = 0.5$\\
        Path Loss Exponent & $\alpha_{L}=2$, $\alpha_{N}=3.5$\\
        Noise Temperature & $T = 290$ K\\
        BS Transmit Power & $P_t = 37$ dBm\\
        Bandwidth & $B = 40$ MHz \\
        Target Data Rate & $R^{tg}_i = 1$ bpc\\
        Maximum RIS Elements & $\mathcal{N}^k = 32 \times 32$\\
        Power Coefficient Vector & $\boldsymbol{\beta} = [0.9895, 0.0101, 0.0003]$\\
        \bottomrule
    \end{tabular}
    }
\end{table}

\begin{table*}[t]
    \centering
    \footnotesize
    \caption{RUOM performance with $\epsilon^{in} = 10^{-1}$ and $\epsilon^{ac} = 10^{-8}$ (UAV 1, UAV 2, UAV 3) $[t = \text{iteration}]$,
    (optimal results in \textbf{bold}).}
    \begin{tabular}{c|c|ccc}
        \toprule
         & $\delta$ & $\lambda = 0.1$ & $\lambda = 0.5$ & $\lambda = 0.9$ \\ 
        \midrule
        \multirow{4}{*}{\rotatebox{90}{$\mathbb{P}_m^{out}$}}
            & \multirow{2}{*}{$10^{-3}$} & $(9.628\mathrm{e}^{-4},4.965\mathrm{e}^{-5},1.370\mathrm{e}^{-10})[t=1]$ & $(9.497\mathrm{e}^{-4},9.676\mathrm{e}^{-4},1.662\mathrm{e}^{-10})[t=1]$ & $(9.173\mathrm{e}^{-4},9.759\mathrm{e}^{-4},9.932\mathrm{e}^{-4})[t=1]$ \\
            & & $(9.628\mathrm{e}^{-4},4.965\mathrm{e}^{-5},1.370\mathrm{e}^{-10})[t=2]$ & $(9.497\mathrm{e}^{-4},6.168\mathrm{e}^{-4},2.045\mathrm{e}^{-10})[t=5]$ & $(9.583\mathrm{e}^{-4},9.583\mathrm{e}^{-4},9.583\mathrm{e}^{-4})[t=38]$ \\
            & \multirow{2}{*}{$10^{-5}$} & $(6.630\mathrm{e}^{-6},8.950\mathrm{e}^{-6},1.370\mathrm{e}^{-10})[t=1]$ & $(7.269\mathrm{e}^{-6},9.694\mathrm{e}^{-6},1.662\mathrm{e}^{-10})[t=1]$ & $(6.778\mathrm{e}^{-6},8.989\mathrm{e}^{-6},9.631\mathrm{e}^{-6})[t=1]$ \\
            & & $(7.110\mathrm{e}^{-6},8.962\mathrm{e}^{-6},1.950\mathrm{e}^{-10})[t=4]$ & $(9.571\mathrm{e}^{-6},9.573\mathrm{e}^{-6},9.573\mathrm{e}^{-6})[t=19]$ & $(9.425\mathrm{e}^{-6},9.425\mathrm{e}^{-6},9.425\mathrm{e}^{-6})[t=96]$ \\
        \midrule
        \multirow{4}{*}{\rotatebox{90}{$N_m^{k^\star}$}}
            & \multirow{2}{*}{$10^{-3}$} & $(38,0,0)[t=1]$ & $(29,30,0)[t=1]$ & $(19,48,59)[t=1]$ \\
            & & $\bold{(38,0,0)^{\star}[t=2]}$ & $\bold{(29,0,0)^{\star}[t=5]}$ & $\bold{(19,0,0)^{\star}[t=38]}$ \\
            & \multirow{2}{*}{$10^{-5}$} & $(61,19,0)[t=1]$ & $(51,109,0)[t=1]$ & $(40,128,178)[t=1]$ \\
            & & $\bold{(61,13,0)^{\star}[t=4]}$ & $\bold{(51,0,0)^{\star}[t=19]}$ & $\bold{(40,69,1)^{\star}[t=96]}$ \\
        \midrule
        \multirow{4}{*}{\rotatebox{90}{$\boldsymbol{\beta}$}}
            & \multirow{2}{*}{$10^{-3}$} & $(0.7111,0.1888,0.0999)[t=1]$ & $(0.7999,0.1062,0.0937)[t=1]$ & $(0.9895,0.0101,0.0003)[t=1]$ \\
            & & $\bold{(0.7111,0.1888,0.0999)^{\star}[t=2]}$ & $\bold{(0.7999,0.1125,0.0874)^{\star}[t=5]}$ & $\bold{(0.9790,0.0204,0.0004)^{\star}[t=38]}$ \\
            & \multirow{2}{*}{$10^{-5}$} & $(0.7111,0.1888,0.0999)[t=1]$ & $(0.7999,0.1062,0.0937)[t=1]$ & $(0.9895,0.0101,0.0003)[t=1]$ \\
            & & $\bold{(0.7099,0.2011,0.0888)^{\star}[t=4]}$ & $\bold{(0.7923,0.2052,0.0024)^{\star}[t=19]}$ & $\bold{(0.9713,0.0263,0.0023)^{\star}[t=96]}$ \\
        \bottomrule
    \end{tabular}
    \label{tab:ruom_performace}
\end{table*}

In Fig. \ref{fig:out_link}, the outage probability is plotted against the number of reflecting elements for RIS-only and composite links, where the outage probability for the direct link corresponds to the zero reflecting elements in the composite link. In the composite link, the outage probability starts from the same point as the direct link, since the power coefficient is fairly allocated among the 3 UAVs using the RUOM algorithm. As the number of reflecting elements increases for RIS-only and composite links, the outage probability decreases at different improvement rates for different UAVs, since the allocated power coefficient is higher for UAV 1 compared to UAV 2 and UAV 2 to UAV 3. Fig. \ref{fig:out_link} also shows that under sufficient power allocation, the RIS-only link for UAV 1 can outperform the composite links of UAV 2 and UAV 3 in terms of outage probability as the number of reflecting elements increases. The initial high outage probability observed in the RIS-only links, especially with a small number of reflecting elements under the best RIS selection $k^{\star}$, is due to severe path loss.

Figs. \ref{fig:out_power} and \ref{fig:out_rate} illustrate the outage probability of 3 UAVs versus the BS transmit power and the target data rate for the composite link, respectively. In Fig. \ref{fig:out_power}, when the BS transmit power increases from 30 dB to 40 dB, the outage probability decreases for the same number of reflecting elements. Also, the speed of improvement increases, leading to enter the second phase of outage probability decrement earlier. Similarly, as the target data rate decreases from 1.5 bpc to 0.7 bpc in Fig. \ref{fig:out_rate}, the outage probability decreases, and it again goes faster to the second phase of decrement. Figs. \ref{fig:out_power} and \ref{fig:out_rate} can also be compared to the composite link in Fig. \ref{fig:out_link} with BS transmit power 37 dB and target data rate 1 bpc. Notably, the power coefficients used in Figs. \ref{fig:out_power} and \ref{fig:out_rate} are the same as Fig. \ref{fig:out_link}.

In Table \ref{tab:ruom_performace}, the performance of the RUOM algorithm is reported under various thresholds $\delta$ and scaling factors $\lambda$. From the initial iteration $[t=1]$ to convergence $[t = t^\star]$, it is clear that both fairness in outage probability and efficiency (reduction) in the number of reflecting elements improve significantly, especially for UAVs 2 and 3. Furthermore, as the scaling factor increases for Case 1 ($\delta=10^{-3}$) and Case 2 ($\delta=10^{-5}$), the outage probability remains below the threshold with improving fairness among UAVs. However, this comes at the cost of more iterations, with a larger portion of power allocated to UAV 1 and a smaller portion to UAV 3. At convergence $[t=t^{\star}]$, the total number of reflecting elements decreases by increasing the scaling factor in Case 1, indicating enhanced efficiency. However, this trend is not valid for Case 2 with the minimum total number of reflecting elements occurring at $\lambda=0.5$. This observation suggests the existence of an optimal scaling factor that balances fairness and efficiency.

\section{Conclusion} \label{conclusion}
In this paper, we introduced a comprehensive analytical framework for RIS-assisted BS-UAV communication in NOMA networks, consisting of three types of links: direct, RIS-only indirect, and composite. For each link type, a closed-form expression for the CDF of SNR under Nakagami-$m$ and double Nakagami-$m$ fading was derived to evaluate the outage performance. Then, we formulated a fairness-efficiency bilevel optimization problem to minimize the maximum outage probability among UAVs while minimizing the total number of required reflecting elements. To solve this, we proposed the iterative RUOM algorithm, which fairly allocates the BS power coefficients while efficiently minimizes the number of RIS reflecting elements. Finally, the simulation results confirmed the accuracy of the analytical models and demonstrated the effectiveness of the RUOM algorithm in achieving fairness and efficiency among UAVs. As a future direction, we will further investigate the fairness-efficiency trade-off under imperfect SIC conditions.

\bibliographystyle{IEEEtran}
\bibliography{ref}

\begin{thebibliography}{10}
\providecommand{\url}[1]{#1}
\csname url@samestyle\endcsname
\providecommand{\newblock}{\relax}
\providecommand{\bibinfo}[2]{#2}
\providecommand{\BIBentrySTDinterwordspacing}{\spaceskip=0pt\relax}
\providecommand{\BIBentryALTinterwordstretchfactor}{4}
\providecommand{\BIBentryALTinterwordspacing}{\spaceskip=\fontdimen2\font plus
\BIBentryALTinterwordstretchfactor\fontdimen3\font minus
  \fontdimen4\font\relax}
\providecommand{\BIBforeignlanguage}[2]{{%
\expandafter\ifx\csname l@#1\endcsname\relax
\typeout{** WARNING: IEEEtran.bst: No hyphenation pattern has been}%
\typeout{** loaded for the language `#1'. Using the pattern for}%
\typeout{** the default language instead.}%
\else
\language=\csname l@#1\endcsname
\fi
#2}}
\providecommand{\BIBdecl}{\relax}
\BIBdecl

\bibitem{FAA_UAV_2025}
\BIBentryALTinterwordspacing
FAA, ``Drones by the numbers (as of 4/1/2025),'' 2025, released: Apr., 2025.
  [Online]. Available: \url{https://www.faa.gov/uas}
\BIBentrySTDinterwordspacing

\bibitem{Badnava-2021-Spectrum}
B.~Badnava, T.~Kim, K.~Cheung, Z.~Ali, and M.~Hashemi, ``Spectrum-aware mobile
  edge computing for {UAV}s using reinforcement learning,'' \emph{2021 IEEE/ACM
  Symposium on Edge Computing (SEC)}, 2021.

\bibitem{Diao-2022-Enhancing}
D.~Diao, B.~Wang, K.~Cao, R.~Dong, and T.~Cheng, ``Enhancing reliability and
  security of {UAV}-enabled {NOMA} comm. with power allocation and aerial
  jamming,'' \emph{IEEE Transactions on Vehicular Technology}, 2022.

\bibitem{Liu-2023-Enabled}
X.~Liu, Y.~Yu, B.~Peng, X.~B. Zhai, Q.~Zhu, and V.~C.~M. Leung, ``{RIS-UAV}
  enabled worst-case downlink secrecy rate maximization for mobile vehicles,''
  \emph{IEEE Transactions on Vehicular Technology}, 2023.

\bibitem{Tang-2025-Throughput}
R.~Tang, J.~Wang, Y.~Zhang, F.~Jiang, X.~Zhang, and J.~Du, ``Throughput max. in
  {NOMA} enhanced {RIS}-assisted multi-{UAV} networks: A deep reinforcement
  learning approach,'' \emph{IEEE Trans. on Vehicular Tech.}, 2025.

\bibitem{Feng-2023-Resource}
W.~Feng, J.~Tang, Q.~Wu, Y.~Fu, X.~Zhang, D.~K.~C. So, and K.-K. Wong,
  ``Resource allocation for power minimization in {RIS}-assisted multi-{UAV}
  networks with {NOMA},'' \emph{IEEE Transactions on Communications}, 2023.

\bibitem{Sobhi-2025-Efficient}
S.~Sobhi-Givi, M.~Nouri, M.~G. Shayesteh, H.~Behroozi, H.~H. Kwon, and M.~J.
  Piran, ``Efficient optimization in {RIS}-assisted {UAV} system using deep
  reinforcement learning for mmwave-{NOMA} {6G} communications,'' \emph{IEEE
  Internet of Things Journal}, 2025.

\bibitem{Zhao-2022-RIS}
J.~Zhao, L.~Yu, K.~Cai, Y.~Zhu, and Z.~Han, ``{RIS}-aided ground-aerial {NOMA}
  communications: A distributionally robust {DRL} approach,'' \emph{IEEE
  Journal on Selected Areas in Communications}, 2022.

\bibitem{Zhao-2025-Exploiting}
S.~Zhao, S.~Gong, B.~Gu, L.~Li, B.~Lyu, D.~Thai~Hoang, and C.~Yi, ``Exploiting
  {NOMA} transmissions in multi-{UAV}-assisted wireless networks: From
  aerial-{RIS} to mode-switching {UAV}s,'' \emph{IEEE Transactions on Wireless
  Communications}, 2025.

\bibitem{FCC_Spectrum_2025}
\BIBentryALTinterwordspacing
FCC, ``Spectrum rules and policies for unmanned aircraft systems,'' 2025,
  released: Feb 11, 2025. [Online]. Available:
  \url{https://www.fcc.gov/document/spectrum-rules-and-policies-unmanned-aircraft-systems}
\BIBentrySTDinterwordspacing

\bibitem{Yang-2020-Performance}
L.~Yang, F.~Meng, J.~Zhang, M.~O. Hasna, and M.~D. Renzo, ``On the performance
  of {RIS}-assisted dual-hop {UAV} communication systems,'' \emph{IEEE
  Transactions on Vehicular Technology}, 2020.

\bibitem{Zhang-2022-Joint}
Q.~Zhang, Y.~Zhao, H.~Li, S.~Hou, and Z.~Song, ``Joint opt. of {STAR-RIS}
  assisted {UAV} comm. systems,'' \emph{IEEE Wireless Comm. Letters}, 2022.

\bibitem{Kim-2019-Impact}
M.~Kim and J.~Lee, ``Impact of an interfering node on unmanned aerial vehicle
  communications,'' \emph{IEEE Trans. on Vehicular Technology}, 2019.

\bibitem{Tegos-2022-Distribution}
S.~A. Tegos, D.~Tyrovolas, P.~D. Diamantoulakis, C.~K. Liaskos, and G.~K.
  Karagiannidis, ``On the distribution of the sum of double-nakagami-$m$ random
  vectors and application in randomly reconfigurable surfaces,'' \emph{IEEE
  Transactions on Vehicular Technology}, 2022.

\bibitem{Channel_Ghazikor_2024}
M.~Ghazikor, K.~Roach, K.~Cheung, and M.~Hashemi, ``Channel-aware distributed
  transmission control and video streaming in {UAV} networks,''
  \emph{arXiv:2408.01885}, 2024.

\bibitem{Abualhayja-2024-Exploiting}
M.~Abualhayja’a, A.~Centeno, L.~Mohjazi, M.~M. Butt, P.~Sehier, and M.~A.
  Imran, ``Exploiting multi-hop {RIS}-assisted {UAV} communications:
  Performance analysis,'' \emph{IEEE Communications Letters}, 2024.

\bibitem{Dorra-2011-SIR}
D.~B. Cheikh, J.-M. Kelif, M.~Coupechoux, and P.~Godlewski, ``{SIR}
  distribution analysis in cellular networks considering the joint impact of
  path-loss, shadowing and fast fading,'' \emph{EURASIP Journal on Wireless
  Communications and Networking}, 2011.

\bibitem{Yang-2022-Performance}
L.~Yang, P.~Li, F.~Meng, and S.~Yu, ``Performance analysis of {RIS}-assisted
  {UAV} comm. systems,'' \emph{IEEE Trans. on Vehicular Tech.}, 2022.

\bibitem{Alqahtani-2021-Performance}
A.~Alqahtani, E.~Alsusa, A.~Al-Dweik, and M.~Al-Jarrah, ``Performance analysis
  for downlink {NOMA} over $\alpha$-$\mu$ generalized fading channels,''
  \emph{IEEE Transactions on Vehicular Technology}, 2021.

\bibitem{Analysis_Ni_2023}
Y.~Ni, H.~Zhao, Y.~Liu, J.~Wang, G.~Gui, and H.~Zhang, ``Analysis of
  {RIS}-aided communications over nakagami-$m$ fading channels,'' \emph{IEEE
  Transactions on Vehicular Technology}, 2023.

\bibitem{Chen-2024-Optimizing}
Z.~Chen, Z.~Ni, P.~Guan, L.~Wang, L.~X. Cai, M.~Hashemi, and Z.~Li,
  ``Optimizing {NOMA} transmissions to advance federated learning in vehicular
  networks,'' \emph{GLOBECOM 2024 - 2024 IEEE Global Communications
  Conference}, 2024.

\bibitem{David-2003-Order}
H.~A. David and H.~N. Nagaraja, \emph{Order Statistics}, 3rd~ed.\hskip 1em plus
  0.5em minus 0.4em\relax Wiley, 2003.

\end{thebibliography}

\end{document}